\documentclass[11pt]{article}
\usepackage{authblk}
\usepackage{amsfonts,amsmath,amssymb,amsthm}
\usepackage{latexsym}
\usepackage{mathrsfs}
\usepackage{tikz}
\usepackage{color}
\usepackage{multirow}

\usepackage[utf8]{inputenc}

\usepackage{amsmath} 
\usepackage{dsfont}
\usepackage[english]{babel}
\usepackage[T1]{fontenc}
\usepackage{physics}
\newcommand{\luc}{\mathfrak{su}(2)}

\makeatletter
\newsavebox{\@brx}
\newcommand{\llangle}[1][]{\savebox{\@brx}{\(\m@th{#1\langle}\)}%
  \mathopen{\copy\@brx\kern-0.5\wd\@brx\usebox{\@brx}}}
\newcommand{\rrangle}[1][]{\savebox{\@brx}{\(\m@th{#1\rangle}\)}%
  \mathclose{\copy\@brx\kern-0.5\wd\@brx\usebox{\@brx}}}
\makeatother

     \numberwithin{equation}{section}
     
\begin{document}

\title{Heun operator of Lie type and the modified algebraic Bethe ansatz }

\author[$\star$]{Pierre-Antoine Bernard}

\author[$\dagger$]{Nicolas Cramp\'e}

\author[$\star$]{Dounia Shaaban Kabakibo}

\author[$\star$]{Luc Vinet}

\affil[$\star$]{Centre de recherches math\'ematiques, Universit\'e de Montr\'eal, P.O. Box 6128, 
Centre-ville Station, Montr\'eal (Qu\'ebec), H3C 3J7, Canada\\

e-mail:bernardpierreantoine@outlook.com, dounia.shaaban.kabakibo@umontreal.ca and vinet@CRM.UMontreal.CA\vspace{5mm}\\ }

\affil[$\dagger$]{Institut Denis-Poisson CNRS/UMR 7013 - Universit\'e de Tours - Universit\'e d'Orl\'eans,
Parc de Grandmont, 37200 Tours, France\\

e-mail: crampe1977@gmail.com}

\maketitle

\vspace{15mm}

\abstract{The generic Heun operator of Lie type is identified as a certain $BC$-Gaudin magnet Hamiltonian in a magnetic field.
By using the modified algebraic Bethe ansatz introduced to diagonalize such Gaudin models, we obtain the spectrum
of the generic Heun operator of Lie type in terms of the Bethe roots of inhomogeneous Bethe equations. 
We show also that these Bethe roots are intimately associated to the roots of polynomial solutions of 
the differential Heun equation.
We illustrate 
the use of this approach in two contexts: the representation theory of $O(3)$ and the computation of the 
entanglement entropy for free Fermions on the Krawtchouk chain.}

\vspace{2cm}
\section{Introduction}

\hspace{5mm} In the spirit of \cite{Bas} where the diagonalization of the Heun--Askey--Wilson operator has been performed by using the modified algebraic Bethe ansatz,
we indicate in the present paper how the Heun operator of Lie type can be diagonalized by a similar method. We show that the particular Heun operator introduced in \cite{CVZ} 
can be identified with the $BC$-Gaudin magnet Hamiltonian in an external magnetic field with one site in a spin $s$ representation of $\mathfrak{su}(2)$.
By a slight generalization of the modified algebraic Bethe ansatz used in \cite{Cra}, we succeed in diagonalizing the generic Heun operator of Lie type.
We then look at two mathematical and physical problems where the Heun operator of Lie type appears and the results apply. 

The algebraic Heun operators are generalizations, introduced in \cite{GVZ}, of the differential Heun operator which is the standardized form of the Fuchsian second order differential
equation with four regular singularities. These operators are the most general bilinear combination of the bispectral pair associated to the orthogonal polynomials of the Askey scheme
and they share the name of these polynomials. The differential Heun operator is recovered in this framework when we consider the dual pair associated to the Jacobi polynomials \cite{GVZ2}.
The Heun--Askey--Wilson operator has been studied in \cite{BTVZ} whereas the Heun--Racah, the Heun--Bannai--Ito and the Heun--Hahn operators have been examined in \cite{BCTVZ,VZ}.
These operators have found applications in different contexts. For example, they give a nice algebraic interpretation of the time-band limiting problem \cite{GVZ} and 
provide the commuting tridiagonal operator \cite{EP} that allows to compute the entanglement entropy for free Fermion models \cite{CNV,CNV2}.
This paper focuses on the simplest algebraic Heun operators which are those of Lie type studied in \cite{CVZ}. In \cite{Bas}, the Heun--Askey--Wilson operator has been identified in the transfer matrix of the XXZ spin chain which allows to use the methods of quantum integrable models to diagonalize this operator. We prove here a similar result: the algebraic Heun operator of Lie type is associated to the $BC$-Gaudin model in an external magnetic field. 
In the spin $s$ representation of $\luc$, this identification allows to use the algebraic Bethe ansatz to diagonalize this Heun operator. 
 
 The usual algebraic Bethe ansatz (or quantum inverse scattering method) has been developed in \cite{STF} to diagonalize integrable model with periodic boundary conditions.
 Its generalization to open boundary conditions has been introduced in \cite{Skl}. 
 The problem of how to apply this method for generic open boundary has remained unsolved for 30 years. 
 In \cite{BC}, the modified algebraic Bethe ansatz was finally introduced to compute the eigenvectors of the XXX spin chain with generic boundaries associated to the eigenvalues found in \cite{CYSW}.
 This modified method has also been used to diagonalize the totally asymmetric exclusion process \cite{CraTASEP} and the XXZ spin chain \cite{BellII, BellIII}
 and it provides the spectrum in terms of inhomogeneous Bethe equations.
 In this paper, we generalize the result of \cite{Cra} where the $BC$-Gaudin model is diagonalized by the modified algebraic Bethe ansatz. An extension is required to take into account the external magnetic field.
 
 The Bethe equations consist in $N$ algebraic relations between $N$ unknowns that need to be solved in order to obtain the eigenvalues and the eigenvectors. These equations usually have multiple solutions and one must demonstrate that all the  eigenvalues are obtained in this manner to prove that the ansatz provides the complete spectrum. 
 One way to tackle the problem is to show that obtaining the solutions of Bethe equations is equivalent to finding the polynomial solutions of a differential (or difference) equation. For a second order differential equation, this problem is referred to the Heine--Stieljes problem and is well-studied.
 This program has already been used with success in \cite{WZ,ISL} for various quantum integrable models.
 We show in this fashion that the spectrum obtained by the modified algebraic Bethe ansatz of the algebraic Heun operator is complete. The associated differential equation
 is an inhomogeneous differential Fuchsian equation with $4$ regular singularities (\textit{i.e.} an inhomogeneous differential Heun operator). 
 We show moreover, by using the Bargmann realization of the Lie algebra $\luc$, that the eigenvalues problem is also associated to a homogeneous differential Fuchsian equation with $5$ regular singularities.  
 
 The plan of the paper is as follows. Section \ref{HL} 
 offers a review of the Heun operators of Lie type \cite{CVZ} and of the algebraic framework needed to study the 
 $BC$-Gaudin magnet. Section \ref{sec:mABA} provides the main steps of the modified algebraic Bethe ansatz for the Gaudin model. In Section \ref{sec:BH}, we study in more details 
 the inhomogeneous Bethe equations and associate their solutions, the Bethe roots, to the roots of a polynomial solution of the differential Heun equation with an inhomogeneous term. 
 Section \ref{sec:hom} focuses on a particular case where the parameters of the algebraic Heun operator satisfy some constraints. It is seen in this case that there exists an alternative to compute the spectrum that 
 leads to homogeneous Bethe equations.
 The last two sections discuss the use of the Heun operator of Lie type in two different contexts: first, in the representation theory of $O(3)$ (see Section \ref{sec:RH}) 
 and second, in the computation of the entanglement entropy for free Fermions on a Krawtchouk chain (see Section \ref{sec:EH}).
 We conclude with an outlook in Section \ref{sec:out}. Useful formulas are gathered in Appendix \ref{app:lw}.

\section{Heun operators and the Gaudin magnet\label{HL}}
 
 \subsection{Algebraic Heun operator of Lie type}

 The Lie algebra $\mathfrak{su}(2)$ is generated by $J_1$, $J_2$ and $J_3$ subject to the following defining relations
 \begin{equation}
  [J_1,J_2]=iJ_3, \qquad  [J_2,J_3]=iJ_1, \qquad  [J_3,J_1]=iJ_2. \label{eq:drsu2}
  \end{equation}
 We set 
 \begin{equation}
  X=\alpha J_1 +\beta J_2, \quad \text{and} \qquad Y=J_1.
 \end{equation}
The Heun operator of Lie type is the following element of the universal enveloping algebra $U(\mathfrak{su}(2))$
\begin{equation}
 W=r_1 [X,Y] + r_2 \{X,Y\} + r_3 X +r_4 Y +r_5, \label{eq:bs}
\end{equation}
where $r_i$ are free parameters and $\{X,Y\}=XY+YX$ is the anti-commutator.
This Heun operator can be rewritten as follows in terms of the Lie generators (up to a renormalization)
 \begin{equation}\label{eq:W}
 W=\rho_1 J_1 + \rho_2 J_2 + \rho_3 J_3 + \{J_1,J_2\} + \rho_4 J_1^2 + \rho_5 
\end{equation}
where $\rho_i$ ($1\leq i \leq 5$) are free real parameters. The normalization of $W$ is chosen so that the coefficient in front of the anti-commutator $\{J_1,J_2\}$ be one.
The parameters $\rho_i$ have been taken real to ensure that $W$ is Hermitian if $J_i$ are represented by Hermitian matrices.  
The generators $X$ and $W$ satisfy the Heun algebra of Lie type studied in \cite{CVZ}.

 \subsection{$BC$-Gaudin magnet in magnetic field}
 
 Let us introduce the $r$-matrix  
\begin{equation}
{r}_{12}(u,v)= \frac{1}{(u-v)(uv-1)}\begin{pmatrix}
       u(1-v^2)&0&0& -2(u-v)\\
       0&-u(1-v^2)& -2v(uv-1)&0\\
       0& -2u(uv-1) & -u(1-v^2) &0\\
        -2uv(u-v)&0&0& u(1-v^2)
      \end{pmatrix}\ 
\label{r12basic}
\end{equation}
solution of the non-standard classical Yang-Baxter equation
 \begin{equation}\label{eq:nsCYBE}
  [\ {r}_{13}(u_1,u_3)\ , \ {r}_{23}(u_2,u_3)\ ]=[\ {r}_{21}(u_2,u_1)\ , \ {r}_{13}(u_1,u_3)\ ]+[\ {r}_{23}(u_2,u_3)\ , \ {r}_{12}(u_1,u_2)\ ]\;,
 \end{equation}
where we denote $r_{12}(u) = r(u)\otimes I\!\! I$ , $r_{23}(u) =I\!\! I\otimes  r(u) $ and so on. 
This $r$-matrix has been used to give the FRT presentations of the Onsager algebra \cite{BBC} and of the classical Askey--Wilson algebra \cite{BC1}.
Let us recall that the parameters $u$ and $v$ are usually referred to as spectral parameters.
This $r$-matrix can be obtained from the one of the $6$-vertex model under a twist by a matrix solution of the classical reflection equation (see \cite{BBC}).\\

Let us now define the $K$-matrix containing as follows the generators of $\mathfrak{su}(2)$
\begin{equation} \label{eq:K}
 K(u)=\frac{2}{(1-a u)(u-a)}\begin{pmatrix}
       u(a^2-1)J_3   &  (a^2u-2a+u)J_1-i u(a^2-1)J_2\\
       -u(a^2-2au+1)J_1+iu(a^2-1)J_2  & -u(a^2-1)J_3 
      \end{pmatrix},
\end{equation}
where $a$ is a free parameter called the inhomogeneity parameter.
The $K$-matrix satisfies the classical reflection equation
\begin{equation}\label{eq:Al}
 [\ K_{1}(u)\ , \ K_{2}(v)\ ]=[\ {r}_{21}(v,u)\ , \ K_{1}(u)\ ]+[\ K_{2}(v)\ , \ {r}_{12}(u,v)\ ].
\end{equation}
In fact, this classical reflection equation is equivalent to the defining relations \eqref{eq:drsu2} of $\mathfrak{su}(2)$.
 
We introduce also a scalar matrix 
\begin{eqnarray}\label{eq:c}
 c(u)&=&\left( \frac{2iu(a^2-1)\rho_1}{a(u^2-1)}  +\frac{2(au-1)(a-u)\rho_2}{a(u^2-1)}  \right)j_1 +    2i\rho_2 j_2 + 2i\rho_3 j_3 +\rho_3 1 \!\!1  ,
\end{eqnarray}
where $j_i$ are the 2 by 2 matrices representing $J_i$ given explicitly by $j_1=\frac{1}{2}\begin{pmatrix}
                                                                                            0 & 1 \\
                                                                                            1 & 0
                                                                                           \end{pmatrix}$, $j_2=\frac{1}{2}\begin{pmatrix}
                                                                                            0 & -i \\
                                                                                            i & 0
                                                                                           \end{pmatrix}   $ and $j_3=\frac{1}{2}\begin{pmatrix}
                                                                                            1 & 0 \\
                                                                                            0 & -1
                                                                                           \end{pmatrix}   $, and $\rho_i$ are the parameters entering in the definition \eqref{eq:W} 
                                                                                           of the Heun operator.
This matrix $c(x)$ is solution of 
\begin{equation}\label{eq:Alc}
0=[\ {r}_{21}(v,u)\ , \ c_{1}(u)\ ]+[\ c_{2}(v)\ , \ {r}_{12}(u,v)\ ].
\end{equation}
Such a scalar solution has been used to study the Gaudin model in a magnetic field (see \cite{Skr}).
It is easy to see that $\check K(x)=K(u) +c(u)$ satisfies also \eqref{eq:Al}.

We introduce at this point the transfer matrix
\begin{equation}
 t(u)= tr (\check K(u)^2 ).
\end{equation}
The important feature of the transfer matrix is its commutativity property for different spectral parameters:
\begin{equation}
 [ t(u), t(v)]=0 .
\end{equation}

We are now in a position to give the first result of this paper: 
amongst the conserved quantities of the Gaudin model (with one site) in a magnetic field, there is the Heun operator of Lie type \eqref{eq:W}.
More explicitly
\begin{equation}
 \frac{a}{8i(1-a^2)u}  t(u)\Big|_{u=0} = W, \label{eq:tw}
\end{equation}
with $\rho_4=\frac{2i(a^2+1)}{a^2-1}$ and $\rho_5=\frac{\rho_1\rho_2}{2}+\frac{i\rho_2^2(1+a^2)}{2(1-a^2)}$. 
Since $\rho_4$ is a real parameter, $a$ must be a pure phase. The parameter $\rho_5$ is given in terms of 
the other parameters but we can add the identity operator to recover the generic Heun operator.

The second result of this paper is the diagonalization of the Heun operator of Lie type $W$ with the Bethe ansatz.

\section{Modified algebraic Bethe ansatz \label{sec:mABA}}

In this section, we recall the results of reference \cite{Cra} and use them to diagonalize the operator $W$ by the algebraic Bethe ansatz.

\paragraph{Gauge transformations.}  Let us introduce the following matrix
\begin{equation}
 M(u)=\begin{pmatrix}
       1/2 & 1/u \\
       -u/2 & 1
      \end{pmatrix}.
\end{equation}
It allows to transform the $r$-matrix \eqref{r12basic}, the $K$-matrix \eqref{eq:K} and the matrix $c(x)$ as follows
\begin{align}
    \widetilde r(u,v) &=M_1(u)^{-1} M_2(v)^{-1} r(u,v) M_1(u) M_2(v) \\
 &=\frac{1}{(uv-1)((u-v)}\begin{pmatrix} -v(u^2-1) & 0 & \frac{2(uv-1)((u-v)}{u}    & 0\\
 0 & v(u^2-1) & -2v(v^2-1) &  -\frac{2(uv-1)((u-v)}{u}  \\
 -\frac{u(uv-1)((u-v)}{2} & -\frac{2u^2(v^2-1)}{v} & v(u^2-1) & 0\\
 0 & \frac{u(uv-1)((u-v)}{2} & 0 & -v(u^2-1)
    \end{pmatrix},
\end{align}
and
\begin{eqnarray}
 &&\widetilde K(u) = M(u)^{-1} K(u) M(u), \quad  \widetilde c(u) = M(u)^{-1} c(u) M(u).
\end{eqnarray}
It is easy to show that the matrices with tilde satisfy relations similar to those obeyed by ${r}(u,v)$, $K(u)$ and $c(u)$. Namely, one gets
\begin{equation}\label{eq:Alt}
 [\ \widetilde K_{1}(u)\ , \  \widetilde K_{2}(v)\ ]=[\  \widetilde{r}_{21}(v,u)\ , \  \widetilde K_{1}(u)\ ]+[\  \widetilde K_{2}(v)\ , \  \widetilde {r}_{12}(u,v)\ ].
\end{equation}
Moreover the gauge transformation does not modify the transfer matrix
\begin{equation}
 t(u)=  tr ( (\widetilde K(u) +  \widetilde c(u)) ^2 ).
\end{equation}
We introduce the new $\mathfrak{su}(2)$ generators: $\widetilde J_1=-\frac{(a^2-1)(a^2-4)}{8a^2} J_1+i\frac{(a^2+1(a^2-4)}{8a^2} J_2+ \frac{a^2+4}{4a}J_3$,
$\widetilde J_2=i\frac{(a^2-1)(a^2+4)}{8a^2} J_1+\frac{(a^2+1)(a^2+4)}{8a^2} J_2 -i\frac{a^2-4}{4a}J_3$ and $\widetilde{J}_3=-\frac{a^2-1}{2a} J_1 +i\frac{a^2-1}{2a}J_2$ so that
\begin{equation}
 \widetilde K(u)= \begin{pmatrix}
                   \frac{2a(u^2-1)}{(au-1)(a-u)}  \widetilde J_3  & \frac{4}{u} \widetilde J_3  + \frac{2(a^2-1)a}{(au-1)(a-u)}\widetilde J_- \\
               -u \widetilde J_3  + \frac{2u^2(a^2-1)}{a(au-1)(a-u)}  \widetilde J_+   &   - \frac{2a(u^2-1)}{(au-1)(a-u)}  \widetilde J_3
                    \end{pmatrix},
\end{equation}
where $\widetilde J_+=\widetilde J_1+i\widetilde J_2$ and $\widetilde J_-=\widetilde J_1-i\widetilde J_2$.
We thus recover a particular case of the $K$-matrix used in \cite{Cra} (see relations (4.1) and (4.2) in \cite{Cra} for $L\to 1$, $x\to u$, $v_1\to a$, $\alpha \to 1$, $\beta \to 0$ and $\gamma \to 0$).

\paragraph{Commutation relations.} As is familiar in the context of algebraic Bethe ansatz, we define the operators $\widetilde A(u)$, $\widetilde B(u)$, $\widetilde C(u)$ and $\widetilde D(u)$ as follows
\begin{equation}
 \widetilde K(u) +  \widetilde c(u)= \begin{pmatrix}
                                     \widetilde A(u) & \widetilde B(u) \\
                                     \widetilde C(u) & \widetilde D(u) 
                                    \end{pmatrix}.
\end{equation}
Only the special case $\widetilde D(u)=-\widetilde A(u)$ has been treated in \cite{Cra}. It does not occur here because of the additional term $\widetilde{c}(u)$. 
We must therefore slightly generalizes the results of \cite{Cra}.
The commutation relations of these operators are computed from relation \eqref{eq:Alt} and  are given by
\begin{align}
 & [\widetilde A(u), \widetilde A(v)] = [\widetilde D(u), \widetilde D(v)] = [\widetilde A(u) ,\widetilde D(v) ]= 0,\\
 & B(u,n)B(v,n+1)=B(v,n)B(u,n+1), \quad C(u,n)C(v,n-1)=C(v,n)C(u,n-1), \label{eq:com2}\\
 &  [\widetilde A(u), B(v,n)]=-[\widetilde D(u), B(v,n)]=\frac{2(u^2-1)}{(u-v)(vu-1)}(vB(v,n)-uB(u,n)),\\
 &  [\widetilde A(u), C(v,n)]=-[\widetilde D(u), C(v,n)]=\frac{2(u^2-1)v^2}{(u-v)(vu-1)}\left(\frac{1}{u}C(u,n)-\frac{1}{v}C(v,n)\right),\\
 & C(u,n)B(v,n)=B(v,n+1)C(u,n+1)+8n \frac{u}{v}\nonumber \\
 & \hspace{3cm}+\frac{2u}{(u-v)(uv-1)v}\left( v(u^2-1)(\widetilde A(u)-\widetilde D(u))-u(v^2-1)(\widetilde A(v)-\widetilde D(v)) \right),
\end{align}
where $B(u,n)=\widetilde B(u)-\frac{2(2n-1)}{u}$ and $C(u,n)=\widetilde C(u)+\frac{(2n-1)u}{2}$.

\paragraph{Shifted transfer matrix.}
Following \cite{Cra}, we define the shifted transfer matrix
\begin{equation}
 t(u,n)= \widetilde A(u)^2 + B(u,n+1)C(u,n+1)+C(u,n)B(u,n) +\widetilde D(u)^2 +2.
\end{equation}
We can see that $t(u,0)=t(u)$ is the transfer matrix we are interested in.
This shifted transfer matrix has the following commutation relation with $B(v,n)$:
\begin{align}
 t(u,n-1) B(v,n) - B(v,n) t(u,n) &= 4 B(v,n) \left( \frac{v(u^2-1)}{(u-v)(uv-1)}\left( \widetilde A(u)-\widetilde D(u) + 4\frac{u^2+1}{u^2-1}   \right) -2 \right)\hspace{1.2cm}\nonumber \\
 &\hspace{-1.7cm} -4 \frac{u}{v}B(u,n)     \left( \frac{u(v^2-1)}{(u-v)(uv-1)}\left( \widetilde A(v)-\widetilde D(v) + 4\frac{v^2+1}{v^2-1}   \right) -4(n-1) \right).
\end{align}

\paragraph{Representation of $\mathfrak{su}(2)$.} 
Denote by $\omega_s$ (with $s \in \mathbb{N}/2$) the highest weight vector of the spin $s$ representation of $\mathfrak{su}(2)$ which satisfies 
\begin{eqnarray}
 \widetilde J_3\, \omega_s =s\, \omega_s,\qquad \widetilde J_+\, \omega_s=0.
\end{eqnarray}
From the explicit form of the operators, one deduces that 
\begin{eqnarray}
&&\hspace{-1.4cm} \widetilde A(u)\omega_s=\alpha(u)\omega_s=\left( \rho_3  +\frac{2as(u^2-1)}{(au-1)(a-u)}  
 -\frac{i(a^2-1)(u^2+1)\rho_1}{2a(u^2-1)} -\frac{(a-u)^2+(au-1)^2}{2a(u^2-1)}\rho_2   \right)\omega_s, \\
&& \hspace{-1.4cm}\widetilde D(u)\omega_s=\delta(u)\omega_s= \left(\rho_3  - \frac{2as(u^2-1)}{(au-1)(a-u)}  
 +\frac{i(a^2-1)(u^2+1)\rho_1}{2a(u^2-1)} +\frac{(a-u)^2+(au-1)^2}{2a(u^2-1)} \rho_2  \right)\omega_s,\\
&&\hspace{-1.4cm}  C(u,n)\omega_s=u \gamma_n\ \omega_s=  u\left( - s +n-\frac{1}{2}-\frac{i(a^2-1)\rho_1}{4a}-\frac{(a^2+1)\rho_2}{4a}+\frac{i\rho_3}{2}   \right)\omega_s. \label{eq:gamma}
\end{eqnarray}

\paragraph{Bethe vectors.} We construct as follows the Bethe vectors that depend on the parameters $\boldsymbol{z}=\{z_1, z_2,\dots z_M\}$:
\begin{equation}
 \mathbb{V}(\boldsymbol{z})=B(z_1,1)B(z_2,2)\dots B(z_M,M) \omega_s.
\end{equation}
Due to relation \eqref{eq:com2}, the entries of the vector $\mathbb{V}(\boldsymbol{z})$ do not depend on the order of the parameters $z_i$.
After some algebraic manipulations, we can express the action of the transfer matrix on the Bethe vector as follows
\begin{eqnarray}
 t(u) \mathbb{V}(\boldsymbol{z})&=& 2 u \gamma_{M+1}  B(z_1,1)B(z_2,2)\dots B(z_M,M)B(u,M+1) \omega_s \nonumber\\
 &&+ \mathcal{W}(u,\boldsymbol{z}) \mathbb{V}(\boldsymbol{z}) \nonumber\\
 &&-\sum_{k=1}^M \frac{16u^2(z_k^2-1)}{(u-z_k)(uz_k-1)z_k}   \mathcal{U}_k(\boldsymbol{z})  \mathbb{V}(\boldsymbol{z}_k,u) \label{eq:to}
\end{eqnarray}
where $\mathbb{V}(\boldsymbol{z}_k,u)=B(z_1,1) \dots B(z_{k-1},k-1) B(u,k) B(z_{k+1},k+1)\dots B(z_M,M) \omega_s$ and 
\begin{eqnarray}
&&\hspace{-1.5cm}\mathcal{W}(u,\boldsymbol{z}) = \lambda(u) +\sum_{k=1}^M \frac{16z_k(u^2-1)}{(u-z_k)(uz_k-1)}\left( \frac{\alpha(u)-\delta(u)}{4} +\frac{u^2+1}{u^2-1}
+\sum_{\genfrac{}{}{0pt}{2}{p=1}{p\neq k}}^{M}\frac{(u^2-1)z_kz_p}{u(z_k-z_p)(z_kz_p-1)}\right)\!,\label{eq:Wa}\\
 &&\hspace{-1cm} \lambda(u)= \alpha^2(u)+\delta^2(u)+2+2u(\alpha'(u)-\delta'(u))+2\frac{u^2+1}{u^2-1}(\alpha(u)-\delta(u)),\\
 &&\hspace{-1cm} \mathcal{U}_k(\boldsymbol{z})=\frac{\alpha(z_k)-\delta(z_k)}{4}+\frac{z_k^2+1}{z^2_k-1}+\sum_{\genfrac{}{}{0pt}{2}{p=1}{p\neq k}}^{M} \frac{z_p(z_k^2-1)}{(z_k-z_p)(z_kz_p-1)}.
\end{eqnarray}
The prime in $\alpha'(u)$ and $\delta'(u)$ stands for the derivative with respect to $u$. 
In the usual algebraic Bethe ansatz, the first line in \eqref{eq:to} is not present. 

\paragraph{Modified algebraic Bethe ansatz and inhomogeneous Bethe equations.} 

For generic values of the parameters, one gets $\gamma_{{M}+1}\neq 0$ for any $M$ ( 
the particular case where there exists a $M$ such that $\gamma_{M}= 0$ is the object of Section \ref{sec:hom}). 
One must compute $B(z_1,1)B(z_2,2)\dots B(z_M,M)B(u,M+1) \omega_s$.
For a generic $M$ there is no simple formula but for $M = 2s$, one gets (see \cite{Cra} for a proof of this result):
\begin{equation}
 2u B(z_1,1)B(z_2,2)\dots B(z_{2s},2s)B(u,2s+1) \omega_s= \overline{\mathcal{W}}(u,\boldsymbol z)\mathbb{V}(\boldsymbol{z}) 
 - \sum_{k=1}^{2s}  \frac{16u^2(z_k^2-1)}{(u-z_k)(uz_k-1)z_k}   \overline{\mathcal{U}}_k(\boldsymbol{z})  \mathbb{V}(\boldsymbol{z}_k,u) 
\end{equation}
where 
\begin{align}
&  \overline{\mathcal{W}}(u,\boldsymbol z) = -8 \gamma^*_{2s+1}  \prod_{p=1}^{2s}\frac{(u-a)(au-1)z_p}{a(u-z_p)(uz_p-1)}, \label{eq:Wab} \\
 &  \overline{\mathcal{U}}_k(\boldsymbol z) = -\, \frac{ \gamma^*_{2s+1}}{2(z_k^2-1)}\ 
 \prod_{ \genfrac{}{}{0pt}{2}{p=1}{p\neq k}  }^{2s}\frac{(z_k-a)(az_k-1)z_p}{a(z_k-z_p)(z_kz_p-1)} ,
\end{align}
and $^*$ is the complex conjugate.
Then, the eigenvalues of the transfer matrix are
\begin{equation}
 {\mathcal{W}}(u,\boldsymbol z)+ \gamma_{2s+1} \overline{\mathcal{W}}(u,\boldsymbol z) ,
\end{equation}
where ${\mathcal{W}}(u,\boldsymbol z)$ is given by \eqref{eq:Wa} with $M=2s$
and $\overline{\mathcal{W}}(u,\boldsymbol z) $ is given by \eqref{eq:Wab}
if  $\boldsymbol{z}$ satisfies the inhomogeneous Bethe equations 
\begin{equation}
\frac{\alpha(z_k)-\delta(z_k)}{4}+\frac{z_k^2+1}{z^2_k-1}+\sum_{\genfrac{}{}{0pt}{2}{p=1}{p\neq k}}^{2s}  \frac{z_p(z_k^2-1)}{(z_k-z_p)(z_kz_p-1)} 
= \frac{|\gamma_{2s+1}|^2}{2(z_k^2-1)}
 \prod_{ \genfrac{}{}{0pt}{2}{p=1}{p\neq k}  }^{2s}\frac{z_p(z_k-a)(az_k-1)}{a(z_k-z_p)(z_kz_p-1)} . \label{eq:bi}
\end{equation}

\section{Bethe roots and Heun operator \label{sec:BH}}

\subsection{Inhomogeneous Heun differential equation}

In this section, we study in more details the inhomogeneous Bethe equations obtained above.

Finding the Bethe roots of the inhomogeneous Bethe equations given by \eqref{eq:bi} is equivalent to computing the roots 
of the monic polynomial solution of degree $2s$ of the following differential Heun equation with an inhomogeneous term 
 \begin{eqnarray}\label{eq:Heuni}
 y''(X) +\left(\frac{a_0}{X}+\frac{a_1}{X-1}+\frac{a_2}{X-A}\right)y'(X)  +\frac{a_3 (X-\mu)}{X(X-1)(X-A)}  y(X)
 = |\gamma_{2s+1}|^2 \frac{(X-A)^{2s}}{X(X-1)},
\end{eqnarray}
where
\begin{align}
 &A=-\frac{(a-1)^2}{4a},\qquad a_0=1 -\frac{i(a^2-1)\rho_1}{4a}-\frac{(a-1)^2\rho_2}{4a}, \qquad a_1=1 -\frac{i(a^2-1)\rho_1}{4a}-\frac{(a+1)^2\rho_2}{4a}, \nonumber\\
 &a_2=-2s, \qquad a_3=|\gamma_{2s+1}|^2 -2s(2s-1+a_0+a_1+a_2)\ . \label{eq:a3}
\end{align}
The roots $Z_i$ of the polynomials $y(X)$ and the Bethe roots $z_i$ are linked by $Z_i=\frac{1}{4}(2-z_i-1/z_i)$.
 The parameter $\mu$ in the previous relation must be chosen so that there exists a polynomial solution. We come back to this point below.

The proof of this statement is quite standard since it is a generalization of the Heine--Stieljes problem. Let us recall the main steps here.
Suppose that for a given $\mu$ there exists a polynomial solution $y(x)$ of degree $2s$ and denote by $Z_i$ the $2s$ roots of this polynomial.
The Heun differential equation \eqref{eq:Heun} at $X=Z_i$ simplifies to
\begin{equation}
 y''(Z_i) +\left(\frac{a_0}{Z_i}+\frac{a_1}{Z_i-1}+\frac{a_2}{Z_i-A}\right)y'(Z_i)=|\gamma_{2s+1}|^2 \frac{(Z_i-A)^{2s}}{Z_i(Z_i-1)}. \label{eq:Hii}
\end{equation}
Defining $z_i$ through $Z_i=\frac{1}{4}(2-z_i-1/z_i)$, one can show that \eqref{eq:Hii} implies that $z_i$ satisfy the Bethe equations.

The inverse is also true. For a given solution $z_i$ of the Bethe equations \eqref{eq:bi}, we define the following monic polynomial of degree $2s$:
\begin{equation}
 y(X)=\prod_{i=1}^{2s}(X-\frac{1}{4}(2-z_i-1/z_i)).
\end{equation}
One can prove that \eqref{eq:Hii} holds for this polynomial
and one deduces that the following polynomial 
\begin{equation}\label{eq:po}
 X(X-1)(X-A)y''(X)+\left( (X-1)(X-A)a_0+a_1 X(X-A)+a_2 X(X-1)\right)y'(X)-|\gamma_{2s+1}|^2(X-A)^{2s+1}
\end{equation}
of degree $2s+1$ vanishes for $X=Z_i=\frac{1}{4}(2-z_i-1/z_i)$ which are the roots of $y(X)$.
Therefore it is equal to $p(X)y(X)$ for some polynomial $p(x)$ of degree 1.
By looking now at the term of degree $2s+1$ in  equation \eqref{eq:po}, we conclude that $p(x)$ is of the form $p(x)=a_3(X-\mu)$ (with $a_3$ given by \eqref{eq:a3}) and that 
$y(X)$ satisfies the inhomogeneous differential Heun equation \eqref{eq:Heuni}.\\

The goal is hence to find $\mu$ such that \eqref{eq:Heuni} has a monic polynomial solution of degree $2s$. 
To arrive at that, we start by putting the polynomial solution 
\begin{equation}\label{eq:yh}
 y(X)=X^{2s}+\sum_{n=0}^{2s-1} c_n X^n
\end{equation}
in the Heun equation \eqref{eq:Heuni}. 
We obtain the following constraints for the coefficients $c_n$, for $0\leq n \leq 2s+1$,
\begin{eqnarray}\label{eq:recuci}
 &&A(n+1)(n+a_0) c_{n+1} - n( (1+A)(a_0+n-1) +a_1 A +a_2)c_n +|\gamma_{n}|^2c_{n-1}\nonumber\\
 &&=\mu a_3 c_n+|\gamma_{2s+1}|^2\begin{pmatrix} 2s+1\\ n \end{pmatrix} (-A)^{2s+1-n}\ ,
\end{eqnarray}
with the conventions $c_{2s}=1$ and $c_{-1}=c_{2s+1}=c_{2s+2}=0$.
The previous relation is directly satisfied for $n=2s+1$ given the explicit value of $a_3$.
Relation \eqref{eq:recuci} for $n=2s$ gives $c_{2s-1}$ as a polynomial of degree 1 with respect to $\mu$.
Then, by recurrence,  relation \eqref{eq:recuci} for $n=2s-p$ ($p=1,\dots 2s-1$) gives $c_{2s-p-1}$ 
as a polynomial of degree $p+1$ with respect to $\mu$. Finally, relation \eqref{eq:recuci} for $n=0$ implies that 
the resulting polynomial $P_{2s+1}(\mu)$ of degree $2s+1$ in $\mu$ must vanish. 

The previous discussion allows to conclude that for each solution $\mu$ of the following relation
\begin{equation}
P_{2s+1}(\mu)=0,
\end{equation}
there exists a polynomial $y_\mu(X)$ solution of the Heun equation \eqref{eq:Heuni} of degree $2s$.
The $2s$ roots of the polynomial $y_\mu(X)$ provide a solution of the Bethe equations \eqref{eq:bi}.
As $P_{2s+1}(\mu)$ is of degree $2s+1$, there are $2s+1$ different $\mu$ and we obtain $2s+1$ different solutions of the Bethe equations.
This proves that the spectrum obtained from the modified algebraic Bethe ansatz is complete.

Let us remark that in the discussion above, we have assumed that $P_{2s+1}(\mu)$ has simple roots. We can expect this to be true for generic parameters.
As far as we know, there is no closed formula for $P_{2s+1}(\mu)$ but it is easy to compute this polynomial from the relations \eqref{eq:recuci}.

\subsection{Eigenvalues of the algebraic Heun operator}

Using the identification \eqref{eq:tw} between the Gaudin model Hamiltonian and the algebraic Heun operator, we can conclude
that the eigenvalues of the algebraic Heun $W$ are given by 
\begin{equation} \label{eq:w1}
 w= 
  \frac{2is|\gamma_{2s}|^2(a-1)}{a+1}-\frac{4ai|\gamma_{2s}|^2}{1-a^2} \sum_{j=1}^{2s} Z_j  
 + \frac{\rho_1\rho_2}{2}+\frac{s(a^2+1)\rho_1-is(a^2-1)\rho_2}{2a} + \frac{i(a^2+1)(4s^2-\rho_2^2)}{2(a^2-1)} 
\end{equation}
where $\{ Z_i\}$ are the roots of a polynomial solution of the inhomogeneous differential Heun equation \eqref{eq:Heuni} 
or equivalently $Z_i=\frac{1}{4}(2-z_i-1/z_i)$ with $\{ z_i\}$ the Bethe roots of \eqref{eq:bi}.

In addition, remarking that $-\sum_{i=1}^{2s} Z_i$ is equal to the coefficient $c_{2s-1}$ in \eqref{eq:yh}, we can express this sum in terms of $\mu$.
Indeed, as explained before, relation \eqref{eq:recuci} for $n=2s$ gives $c_{2s-1}$ as a polynomial in $\mu$ of degree 1. 
Therefore, we can conclude that 
\begin{equation}
 w=\frac{4ai a_3\mu}{1-a^2} +\rho_1(\frac{\rho_2}{2}-s)-\frac{i(a^2+1)(\rho_2^2-2s(s+1))}{2(a^2-1)}+\frac{i(a-1)(\rho_2-|\gamma_{2s+1}|^2)}{(a+1)} +\frac{2is(s-1)a}{a^2-1}.
\end{equation}
This relation allows to view $w$, the eigenvalues of the algebraic Heun operator, as the parameters in the differential inhomogeneous Heun operator \eqref{eq:Heuni}
such that this equation has a polynomial solution. Similar results can be obtained directly from the Bargmann representation of $\luc$ as explained in the following subsection.

\subsection{Bargmann realization of the algebraic Heun operator of Lie type \label{sec:bar}}

It is well-known that the spin $s$ representation of $\luc$ can be realized in terms of differential operators acting on the space of univariate polynomials of order less or equal to $2s$:
\begin{align}
&J_1=\frac{1-z^2}{2}\frac{d}{dz}+sz,\qquad J_2=\frac{1+z^2}{2i}\frac{d}{dz}-\frac{sz}{i},\qquad J_3=s-z\frac{d}{dz}.
\end{align}
Indeed, the actions on the monomials
\begin{align}
&\lvert s,m \rangle=\frac {z^{s-m}}{ \sqrt{ (s-m)!(s+m)!}}, \quad\text{for } -s\leq m \leq s,
\end{align}
are given by
\begin{align}
&J_{\pm}\lvert s,m \rangle=\left ( J_1\pm iJ_2\right )\lvert s,m \rangle= \sqrt{(s\mp m)(s\pm m+1)}\lvert s,m\pm1 \rangle,\qquad J_3\lvert s,m \rangle=m\lvert s,m \rangle,
\end{align}
which is the standard spin $s$ representation of $\luc$.

The algebraic Heun operator $W$ in this realization becomes: 
\begin{align}
 \nonumber W&=\frac{i(z^2-1)(z^2a^2-1)}{(a^2-1)}\frac{d^2}{dz^2}+
  \bigg( \frac{1-2s}{2}(z^3\left( \rho_4+2i \right)-\rho_4z)-z^2 \frac{\rho_1+i \rho_2}{2}-z\rho_3+\frac{\rho_1-i \rho_2}{2}
\bigg)\frac{d}{dz}\hspace{5mm} \\
&+z^2 \left(s^2 \rho_4+2 i s^2-\frac{s \rho_4}{2}-i s\right)+s z (\rho_1+i \rho_2)+s \rho_3+\frac{s \rho_4}{2}+\rho_5 \label{Wbarg}
\end{align}
with $\rho_4=\frac{2i(a^2+1)}{a^2-1}$ and $\rho_5=\frac{\rho_1\rho_2}{2}+\frac{i\rho_2^2(1+a^2)}{2(1-a^2)}$. 
Then, the eigenvalues $w$ of the algebraic Heun operator $W$ can be seen as the eigenvalues of this differential operator
\begin{equation}\label{eq:ev}
 W \phi(z)= w \phi(z).
\end{equation}
Thus, $w$ is obtained by asking that there is a polynomial solution $\phi(z)$ of degree less or equal to $2s$ of this differential equation. 
This operator $W$ leads to a Fuchsian second order
differential equation with five regular singularities $\{1,-1,1/a,-1/a,\infty\}$, but in the special case where $\rho_1=\rho_2=0$, it reduces to the differential Heun operator.
Indeed, let us perform the change of variable
\begin{equation}
 y=a^2z^2 \quad \text{and} \qquad \psi(y) = \phi( a^2 z^2 ) .
\end{equation}
In this new variable when $\rho_1=\rho_2=0$, \eqref{eq:ev} becomes the Heun differential equation
\begin{align}
&\frac{d^2\psi}{dy^2}+\bigg (\frac { 1}{2y}+\frac{1-2s-i\rho_3}{2(y-1)}+\frac{1-2s+i\rho_3}{2(y-a^2)}\bigg )\frac{d\psi}{dy}\nonumber\\ &+\frac{2(2s-1)sy+is\rho_3(1-a^2)+s(a^2+1)}{4y(y-1)(y-a^2)}\psi=\frac{iw(1-a^2)}{4y(y-1)(y-a^2)}\psi. \label{eq:db2}
\end{align}
Comparing this differential Heun operator
obtained from the Bargmann realisation and the one in \eqref{eq:Heuni} obtained from the Bethe ansatz, we see that one singularity of the former is located at $a^2$ whereas for the latter one singularity is at $A=-\frac{(a-1)^2}{4a}$.

\section{Homogeneous case\label{sec:hom}}

In the previous sections, we have explained how the modified algebraic Bethe ansatz can be used to diagonalize the transfer matrix of the Gaudin model and the Heun operator, that leads to inhomogeneous Bethe equations.
However, for particular values of the parameters, there exist homogeneous Bethe equations which diagonalize also this model.
Indeed, let us suppose that $s$, $a$, $\rho_1$, $\rho_2$ and $\rho_3$ are such that there is an integer 
$0\leq \mathcal M \leq 2s-1$ satisfying $\gamma_{\mathcal{M}+1}=0$ \textit{i.e.}
\begin{equation}
\begin{cases}
\displaystyle \mathcal{M}= s  -\frac{1}{2}+\frac{i(a^2-1)\rho_1}{4a}+\frac{(a^2+1)\rho_2}{4a}   \\
 \rho_3=0
 \end{cases},\label{eq:cM}
 \end{equation}
where the real parameter $\rho_3$ is set to zero to ensure that the imaginary part of the relation $\gamma_{\mathcal{M}+1}=0$ is verified.
In this case, some eigenvalues of the transfer matrix are given by $\mathcal{W}(u,\boldsymbol{z})$ (relation \eqref{eq:Wa} for $M=\mathcal{M}$),
if $\boldsymbol{z}$ satisfies the Bethe equations $\mathcal{U}_k(\boldsymbol{z})=0$ \textit{i.e.}
\begin{equation}\label{eq:bh}
 \frac{1}{4}(\alpha(z_k)-\delta(z_k))+\frac{z_k^2+1}{z^2_k-1}+\sum_{\genfrac{}{}{0pt}{2}{p=1}{p\neq k}}^{\mathcal{M}} \frac{z_p(z_k^2-1)}{(z_k-z_p)(z_kz_p-1)} =0.
\end{equation}
In this case, the Bethe equations are called homogeneous.
As before, finding the Bethe roots $z_k$, solutions of the Bethe equations \eqref{eq:bh} is the same as finding
polynomial solutions of a differential equation.
More precisely, it amounts to finding polynomial solutions of degree $\mathcal{M}$ of the homogeneous Heun differential equation
\begin{equation}\label{eq:Heun}
 y''(X) +\left(\frac{a_0}{X}+\frac{a_1}{X-1}+\frac{a_2}{X-A}\right)y'(X)  +\mathcal{M}^2\, \frac{ (X-\mu)}{X(X-1)(X-A)}  y(X)=0,
\end{equation}
for a suitable $\mu$, where 
\begin{align}
 &A=-\frac{(a-1)^2}{4a},\qquad a_0=s+\frac{1}{2}-\mathcal{M}+\frac{\rho_2}{2}, \qquad a_1=s+\frac{1}{2}-\mathcal{M}-\frac{\rho_2}{2}, \qquad a_2=-2s.
 \end{align}
If $\{Z_i\}$ are the $\mathcal{M}$ roots of a polynomial solution of the Heun differential equation, then, for each $Z_i$, take $z_i$ as one solution of $Z_i=\frac{1}{4}(2-z_i-1/z_i)$, then these $\{z_i\}$ are solutions of the Bethe equations \eqref{eq:bh}.

The parameter $\mu$ must be chosen such that \eqref{eq:Heun} has a polynomial solution of degree $\mathcal{M}$. 
If we take the polynomial solution of the form
\begin{equation}\label{eq:yhh}
 y(X)= X^\mathcal{M}+ \sum_{n=0}^{\mathcal{M}-1} c_n X^n
\end{equation}
upon inserting this expression in the Heun equation  \eqref{eq:Heun}, we obtain the following constraints for the coefficients $c_n$, for $0\leq n \leq \mathcal M+1$,
\begin{equation}\label{eq:recuc}
 A(n+1)(n+a_0) c_{n+1} - n( (1+A)(a_0+n-1) +a_1 A +a_2)c_n +(\mathcal{M}+1-n)^2c_{n-1}=\mu \mathcal{M}^2 c_n,
\end{equation}
with the conventions $c_\mathcal{M}=1$ and $c_{-1}=c_{\mathcal{M}+1}=c_{\mathcal{M}+2}=0$.
Again, these equations are consistent only if $\mu$ is a root of the polynomial $P_{\mathcal{M}+1}(\mu)$ of degree $\mathcal{M}+1$ in $\mu$.

When the constraints \eqref{eq:cM} are satisfied, the eigenvalues of the algebraic Heun operator $W$ \eqref{eq:W} 
can also be deduced from those of the Gaudin transfer matrix thanks to formula \eqref{eq:tw}, they are given by
\begin{equation}
 w=  \frac{i}{a^2-1} \left( s(a^2+1)-2\mathcal{M}a-a\rho_2 \right)+4\frac{ia}{a^2-1} \sum_{i=1}^\mathcal{M} Z_i\ ,
\end{equation}
where $\{ Z_i\}$ are the roots of a polynomial solution of the differential Heun equation \eqref{eq:Heun} 
or equivalently $Z_i=\frac{1}{4}(2-z_i-1/z_i)$ with $\{ z_i\}$ the Bethe roots of \eqref{eq:bh}.

Since $-\sum_{i=1}^\mathcal{M} Z_i$ is equal to the coefficient $c_{\mathcal{M}-1}$ in \eqref{eq:yh}, we can express this sum in terms of $\mu$
as follows
\begin{equation}
 i \sum_{i=1}^\mathcal{M} Z_i=-ic_{\mathcal{M}-1} =-i(\mu \mathcal{M}^2   +\mathcal{M}( (1+A)(a_0+\mathcal{M}-1) +a_1 A +a_2))  \ . 
\end{equation}
Then, the eigenvalues of the algebraic Heun operator $W$ \eqref{eq:W} become
\begin{equation}
 w= \frac{4ia\mu \mathcal{M}^2}{1-a^2} + \frac{i}{1-a^2} \left(  ( (a^2+1)s +a\rho_2)(2\mathcal{M}+1)-\mathcal{M}^2(a-1)^2 \right),
\end{equation}
where $\mu$ are the roots of the polynomials $P_{\mathcal{M}+1}(\mu)$.

The previous construction provides $\mathcal{M}+1$ eigenvalues. They belong to the spin range $\{s-\mathcal{M},s-\mathcal{M}+1,\dots, s\}$.
We remark that this spin domain is stabilized by the algebraic Heun operator since (when relations \eqref{eq:cM} hold)  
\begin{equation}
 W_{\mathcal{M}+1,\mathcal{M}+2}W_{\mathcal{M}+2,\mathcal{M}+1}=0.
\end{equation}

To obtain the second part of the spectrum, we must start from the Bethe vectors $\overline{\mathbb{V}}(\overline{\boldsymbol{z}})$ constructed from the lowest weight of $\mathfrak{su}(2)$.
We give the definition and the useful formulas in Appendix \ref{app:lw}. We note that for $M=2s-1-\mathcal{M}$, one gets $\beta_{-M}=0$ and relation \eqref{eq:tob} simplifies since the first line vanishes.
Therefore, the usual Bethe ansatz works and one finds that $\overline{\mathcal{W}}(u,\overline{\boldsymbol{z}})$ (see \eqref{eq:Wabb}) for $M=2s-1-\mathcal{M}$ is an eigenvalue if 
 $\overline{\boldsymbol{z}}$ satisfies the Bethe equations, 
\begin{equation}
 \frac{\overline{\delta}(z_k)-\overline{\alpha}(z_k)}{4}
  +\frac{\overline{z}_k^2+1}{\overline{z}^2_k-1}
  +\sum_{\genfrac{}{}{0pt}{2}{p=1}{p\neq k}}^{2s-1-\mathcal{M}} \frac{\overline{z}_p(\overline{z}_k^2-1)}{(\overline{z}_k-\overline{z}_p)(\overline{z}_k\overline{z}_p-1)}=0, \label{eq:hb}
\end{equation}
for $1\leq k \leq 2s-1-\mathcal{M}$. Once more, finding the Bethe roots $\overline{\boldsymbol{z}}$ is equivalent to finding a polynomial solution of a differential Heun equation.
In this way, we prove that we obtain $2s-\mathcal{M}$ solutions for spins running from $-s$ to $s-\mathcal{M}-1$ in unit steps.

From the Bethe vectors $\mathbb{V}(\boldsymbol{z})$ and $ \overline{\mathbb{V}}(\overline{\boldsymbol{z}})$, 
we obtain the complete spectrum of the Gaudin model or of the algebraic Heun operators when relations \eqref{eq:cM} hold.

\section{Representations of the rotation group $O(3)$ and Heun operator\label{sec:RH}} 

A special and important case of the Heun operator of Lie type arises when $\rho_1=\rho_2=\rho_3=0$:
\begin {equation}
W|_{\rho_1=\rho_2=\rho_3=0}= \{J_1,J_2\} +  \frac{2i(a^2+1)}{a^2-1} J_1^2 .
\end{equation}
It corresponds to the situation where there is no external magnetic field for the Gaudin magnet (\textit{i.e.} the scalar matrix $c(u)$ given in \eqref{eq:c} vanishes).
Consider the following rotation of the  generators of $\luc$
\begin{eqnarray}
 J_1=\cos(\theta)\bar{J_1}-\sin(\theta)\bar{J_2},\quad\text{and} \qquad  J_2=\sin(\theta)\bar{J_1}+\cos(\theta)\bar{J_2},
 \end{eqnarray}
 with $a=e^{2i\theta}$.
 In terms of the generators $\bar{J_1}$ and $\bar{J_2}$, the algebraic Heun operator reads 
\begin{align}
&E:= \frac{4i(1-a)}{1+a}W|_{\rho_1=\rho_2=\rho_3=0}=4(\bar{J_1}^2+r\bar{J_2}^2),\qquad r=\left(\frac{1-a}{1+a}\right)^2.
 \end{align} 
 
The operator $E$ occurs in many physical and mathematical contexts.
It is seen to be equivalent (up to an affine transformation) to the Hamiltonian of the quantum Euler top \cite{T}.
It also appears in the representation theory of the group $O(3)$ and its universal covering group $SU(2)$ as follows. 
As is very familiar, the standard representation basis is defined by the joint eigenvectors of the Casimir element $C=J_1^2+J_2^2+J_3^2$ and of the generator $J_3$. It is said to be of subgroup type since it corresponds to the group reduction $O(3) \supset O(2)$ with one subgroup generator, $J_3$, diagonalized.
Close to 50 years ago, Patera and Winternitz stressed \cite{PW_Heun} the existence of a second interesting basis stemming from the classification of second order polynomials in the generators.
This second basis is provided by the eigenfunctions of $C$ and $E=4(J_1^2+rJ_2^2)$
(and of the discrete operators $X$ and $PZ$, where $X$ and $Z$ correspond to reflections in the $yz$ and $xy$ planes and $P$ is the parity operator). It is not of subgroup type as $E$ is not the generator of any subgroup of $O(3)$. 
Ways to obtain the eigenvalues of E are described in \cite{PW_Heun} but none are providing a closed formula. Our construction via the Bethe ansatz strikingly advances the characterization of the eigenvalues $\epsilon$ of $E$. Indeed, by specializing formulas \eqref{eq:bi} and \eqref{eq:w1}, one gets that 
the eigenvalues of $E$ are given by
\begin{equation} \label{eq:e1}
 \epsilon=\frac{2s(4s^2+1)(a^2+1)}{(a+1)^2}-\frac{4a(s-\frac{1}{2})^2}{(a+1)^2}\sum_{p=1}^{2s}(z_p+z_p^{-1}),
\end{equation}
with the Bethe roots satisfying
\begin{equation}
\frac{as(z_k^2-1)}{(az_k-1)(a-z_k)}+\frac{z_k^2+1}{z^2_k-1}+\sum_{\genfrac{}{}{0pt}{2}{p=1}{p\neq k}}^{2s}  \frac{z_p(z_k^2-1)}{(z_k-z_p)(z_kz_p-1)} 
= \frac{(2s+1)^2}{8(z_k^2-1)}
 \prod_{ \genfrac{}{}{0pt}{2}{p=1}{p\neq k}  }^{2s}\frac{z_p(z_k-a)(az_k-1)}{a(z_k-z_p)(z_kz_p-1)} . 
\end{equation}
Let us also remark that if $s$ is a half-integer, the result of Section \ref{sec:hom} can also be used to characterize the spectrum of $E$. In this case the eigenvalues read
\begin{equation} \label{eq:e2}
 \epsilon= \frac{4s(a^2+1)}{(a+1)^2}-\frac{4a}{(a+1)^2}\sum_{p=1}^{s-\frac{1}{2}}(z_p+z_p^{-1}),
\end{equation}
with the Bethe roots satisfying
\begin{equation}\label{eq:bhd}
\frac{as(z_k^2-1)}{(az_k-1)(a-z_k)}+\frac{z_k^2+1}{z^2_k-1}+\sum_{\genfrac{}{}{0pt}{2}{p=1}{p\neq k}}^{s-1/2} \frac{z_p(z_k^2-1)}{(z_k-z_p)(z_kz_p-1)} =0.
\end{equation}
The specialization used here of relation \eqref{eq:db2} for the Bargmann realization of the Heun operator reproduces  the result of \cite{PW_Heun}.
We should point out that the supplementary condition $0<r<1$ is imposed in \cite{PW_Heun} which implies that $a=\frac{1-\sqrt{r}}{1+\sqrt{r}}$. In the construction above we have required that
$a$ be a pure phase but the results can easily be generalized and the relations \eqref{eq:e1}-\eqref{eq:bhd} are still valid for any $a$.  

\section{Entanglement entropy for the Krawtchouk chain  and Heun operator \label{sec:EH}}

In this section, we present another interesting problem to which the results of this paper can be applied, namely the determination of the entanglement entropy for free Fermions on Krawtchouk chains. These chains are interesting in many respects. They emerge from the projection of spin systems defined on hypercubes \cite{CDD} and they are known to allow for perfect state transfer \cite{ACDE}. They are usually introduced in the following way. Let us consider the free Fermion inhomogeneous Hamiltonian with nearest neighbor interactions and local magnetic fields defined by:
\begin{align}
    \hat{\mathcal{H}} = \frac{\beta}{2} \sum_{n=0}^{2s-1} \sqrt{(n+1)(2s - n)}( c_n^\dagger c_{n+1} + c_{n+1}^\dagger c_n) - \alpha \sum_{n=0}^{2s}( n - s) c_n^\dagger c_{n},
    \label{hamil}
\end{align}
\noindent where $\{c_m^\dagger, c_n\} = \delta_{mn}$ and where $\alpha$ and $\beta$ are free parameters. We can choose the normalization of $\mathcal{H}$ such that $\alpha^2 + \beta^2 = 1$ and take $\alpha = \cos{(2\theta)}$ and $\beta = \sin{(2\theta)}$. In what follows, it will be more convenient to rewrite \eqref{hamil} in terms of the matrices $J_1$ and $J_3$ from an irreducible representation of $\mathfrak{su}(2)$ of dimension $2s + 1$. Indeed, we have that
\begin{align}
    \hat{\mathcal{H}} = 
\begin{pmatrix}
c_0^\dagger & \dots & c_{2s}^\dagger
\end{pmatrix}
\hat{H}
\begin{pmatrix}
c_0\\
\vdots\\
c_{2s}
\end{pmatrix},
\end{align}
\noindent where
\begin{align}
    \hat{H} = \cos{(2\theta)} J_3 + \sin{(2\theta)} J_1.
\end{align}
\noindent We easily find an orthonormal basis $\{\ket{\omega_k}\}_{k \in \{0, \dots, 2s\}}$ of $\mathbb{C}^{2s+1}$ such that  $\hat{H}\ket{\omega_k} = \omega_k \ket{\omega_k}$. The eigenvalues are given by $\omega_k = k - s$ and the diagonalized Hamiltonian is
\begin{align*}
    \hat{\mathcal{H}} = \sum_{k = 0}^{2s} \omega_k \tilde{c}_k^\dagger \tilde{c}_k,
\end{align*}
\noindent where the $\tilde{c}_k = \sum_{n = 0}^{2s} \bra{n}\ket{\omega_k} c_n$  also respect $\{\tilde{c_j}, \tilde{c_k}^\dagger\} = \delta_{jk}$. The overlap coefficients are given in terms of Krawtchouk polynomials \cite{KS}:
\begin{equation}
    \begin{split}
          \bra{n}\ket{\omega_k} = \sqrt{\binom{2s}{n}\binom{2s}{2s - k}} (\sin{\theta})^{2s}\ |\cot{\theta}|^{k-n} K_n(2s-k; \sin^2{\theta}, 2s).
    \end{split}
    \label{overl}
\end{equation}

The ground state $|\Psi_0 \rrangle$ is defined by filling up the Fermi sea. If $|0 \rrangle$ represents the vacuum state which is annihilated by all the $\tilde{c}_k$, we have
\begin{align}
    |\Psi_0 \rrangle = \tilde{c}^\dagger_K \tilde{c}^\dagger_{K-1} \dots \tilde{c}^\dagger_1 \tilde{c}^\dagger_0\  |0 \rrangle.
\end{align}
\noindent $K$ is taken to be the largest integer $k$ such that states associated to a negative energy $\omega_k$ are filled.  A natural question to ask about such chain is the following: if we split it in two, what is the entanglement between the two parts? This information is contained in the entanglement entropy:
\begin{align}
    S = \text{tr}(\rho \ln{(\rho)}),
\end{align}
\noindent where $\rho$ is the reduced density matrix associated to one of the two parts. Here, we take the subsystem to be the first $l+1$ sites of the chain. In particular, the projection operator $\pi_\ell$ over the subsystem is taken to be
\begin{align}
    \pi_\ell = \sum_{n = 0}^\ell \ket{n}\bra{n}.
\end{align}

For a Krawtchouk chain in its ground state $|\Psi_0\rrangle$, it is known that the value of $S$ can also be extracted from the eigenvalues of the chopped correlation matrix \cite{CNV2}. This matrix is obtained by first considering the complete correlation matrix, which is the $(2s + 1) \times (2s+1)$ matrix having for entries:
\begin{align}
    \hat{C}_{mn} = \llangle \Psi_0 | c_m^\dagger c_n |\Psi_0 \rrangle.
\end{align}
\noindent It is useful to note that $\hat{C}$ can be expressed as:
\begin{align}
    \hat{C} =  \sum_{k = 0}^K\ket{\omega_k}\bra{\omega_k}.
    \label{cc}
\end{align}
\noindent Then, the chopped correlation matrix associated to the first $\ell + 1$ sites in the chain is obtained by considering the submatrix of $\hat{C}$ defined as:
\begin{align}
    C = \pi_\ell\  \hat{C}\  \pi_\ell = |\hat{C}_{mn}|_{0 \leq m,n \leq \ell}.
\end{align}
\noindent Given \eqref{overl}, we see that
\begin{align}
    C_{mn} = \sum_{k=0}^K \sqrt{\binom{2s}{n}\binom{2s}{m}}\binom{2s}{k}(\sin^2{\theta})^{2s}|\cot{\theta}|^{\tiny{2k - n+m}} K_m(2s-k) K_n(2s-k),
\end{align}
\noindent where the last two parameters of the Krawtchouk polynomials are kept implicit. 

In general, we see that the submatrices of \eqref{cc} have a rather complicated expression and do not prove easy to diagonalize.  However, considerations of bispectrality in the context of the time and band limiting problem have shown how to identify a tridiagonal operator $T$, in fact an algebraic Heun operator \cite{GVZ,CNV2}, that has the property of commuting with the chopped correlation matrix. It is expressed as 
\begin{align}
    T = \{\hat{H}, J_3\} + \mu J_3 + \nu \hat{H},
\end{align}
\noindent where $\mu = -2K - 1 + 2s$ and $\nu = 2\ell + 1 - 2s$ \cite{CNV}. Thus, one could instead diagonalize $T$ and use the results to extract the spectrum of $C$. This is where the modified algebraic Bethe ansatz comes into play. If we make the identification 
\begin{align}
    {-\sin{(2\theta)}}J_1 +{\cos{(2\theta)}}J_3  \rightarrow \tilde{J}_1, \quad  {\cos{(2\theta)}}J_1 +{\sin{(2\theta)}}J_3  \rightarrow \tilde{J}_2
\end{align}
\noindent for the generators and
\begin{equation}
    \begin{split}
        &\rho_1 = {\cot{(2\theta)}\mu  + \frac{\nu}{\sin{(2\theta)}}}, \quad \quad  \rho_2 = {\mu }, \quad \quad  \rho_3 = 0,\\
         &\rho_4 = 2\cot{2\theta} \quad \quad \text{and} \quad \quad \rho_5 = \frac{\mu \nu}{2\sin{2\theta}} 
    \end{split}
\end{equation}
\noindent for the parameters, we see that
\begin{align}
    T = \sin{2\theta} (W - \rho_5) ,
\end{align}
\noindent where $W$ is defined in \eqref{eq:bs}. We thus see that the spectrum of $T$ is given, up to an affine transformation, by the one of $W$.\\

Moreover, it is observed that this $W$ belongs to the homogeneous case (discussed in Section 5). Indeed, one can check that we here have $\mathcal{M} = \ell $ when considering the Bethe vector constructed with the lowest weight. Therefore, the $\ell + 1$ eigenvalues $t_{\boldsymbol{\Bar{z}}}$ of $T$ will be given by:
\begin{equation}
 \ t_{\boldsymbol{\Bar{z}}} =  s\cos{(2 \theta)} + \frac{\mu}{2} - \frac{1}{2} \sum_{i=1}^\mathcal{\ell} (\Bar{z}_i+ \frac{1}{\Bar{z}_i}),
\end{equation}
\noindent where the $\boldsymbol{\Bar{z}}$ are solutions to the Bethe equations given by \eqref{eq:hb}:  
\begin{equation}
\begin{split}
   \frac{e^{2i\theta} s (\Bar{z}_k^2 - 1)}{(e^{2i\theta} \Bar{z}_k - 1)(e^{2i\theta} - \Bar{z}_k)} + \frac{(\Bar{z}_k^2 +1)(1 - \frac{\nu}{2})- \mu \Bar{z}_k}{(\Bar{z}_k^2 - 1)  } +\sum_{\genfrac{}{}{0pt}{2}{p=1}{p\neq k}}^{l} \frac{\Bar{z}_p(\Bar{z}_k^2-1)}{(\Bar{z}_k-\Bar{z}_p)(\Bar{z}_k \Bar{z}_p-1)} =0.
\end{split}
 \label{BeqC} 
\end{equation}
\noindent At this point, it is possible to recover the spectrum of $C$. Usually, this is done by acting with $C$ on the eigenvectors of $T$, \textit{i.e.} the Bethe vectors for $W$ \cite{CKV}. Since these are also eigenvectors of the chopped correlation matrix, we can read the eigenvalues of $C$ from the results. We here want to present an alternative way of obtaining the spectrum. It amounts to constructing a polynomial $P$ of order $\ell$ such that
\begin{align}
    C = P(T) = \sum_{j=0}^\ell a_j T^j.
\end{align}
\noindent If we have $P$, then it is easy to see that $P(t_{\boldsymbol{z}})$ gives the eigenvalue of $C$ associated to $\boldsymbol{z}$. To prove the existence of $P$, it is sufficient to know that $[C,T]=0$ is verified and to notice that, since $T$ is tridiagonal, the first $\ell+1$ powers of $T$ are linearly independent. The fact that $T$ is tridiagonal also allows to determine $P$. Indeed, since we have that
\begin{align}
    \bra{0} T^r \ket{m} = 0 \quad \quad \text{if } r < m,
\end{align}
\noindent we are led to
\begin{align}
 a_{\ell} = \bra{0}C\ket{\ell}/\bra{0}T^\ell\ket{\ell}
\end{align}
\noindent and to the recurrence relation:
\begin{align}
    a_{\ell - j} = \frac{
    \Big[\bra{0}C\ket{\ell-j} - \sum_{r = 0}^{j-1} a_{\ell - r} \bra{0}T^{\ell-r}\ket{\ell - j}\Big]}{\bra{0}T^{\ell-j}\ket{\ell - j}}.
\end{align}
\noindent Given this relation, constructing $P$ is straightforward and obtaining the eigenvalues of $C$ and the entanglement entropy follows.

\section{Conclusion and outlook \label{sec:out}} 

 This paper has woven threads between the algebraic Heun operator of Lie type, the differential Heun operator, a Fuchsian second order
differential equation with five regular singularities, the inhomogeneous Bethe equation and the Gaudin magnet. It showed in particular that there exist two equivalent ways to compute the eigenvalues of the algebraic Heun operator of Lie type by studying the
polynomial solution of two differential equations, one inhomogeneous with four singularities and the other
homogeneous with five singularities.
This type of equivalence has already been noticed in a different context \cite{CN} and certainly deserves further investigation.

We have examined the Gaudin model with only one site. It
is well-known however that the model with $N$ sites remains integrable;
the Bethe equations are also given in \cite{Cra}. In the framework of the Bargmann realization, a differential equation with $N$ variables is obtained in that case. A natural question is to compare this differential equation with the one obtained from the Bethe equations. This would provide a generalization to $N$ variables of the results of this paper.

We showed that the tools of quantum integrable systems and the algebraic
Heun operators of Lie type are relevant to the representation theory of $\luc$. We expect that other quantum integrable systems and algebraic Heun operators of different types will find their way in the representation theory of higher rank Lie algebras.
For example, algebraic Heun--Hahn or Racah--Hahn operators appear in \cite{CPV} in the study of the diagonal centralizer of two
$\mathfrak{su}(3)$.

Finally, we demonstrated the usefulness of the Bethe ansatz to compute the entanglement entropy of free Fermion chain with couplings given by the recurrence coefficients of the Krawtchouk polynomials. It would be interesting to use this approach for other chains.
Algebraic Heun operators of different type which commute with the chopped correlation matrix associated to different inhomogeneous free Fermion chains have been introduced in \cite{CNV, CNV2}. The computation of their spectra with the Bethe ansatz could be used to determine the entanglement entropy at least in the thermodynamical limit. We plan on returning to some of these questions.
\vspace{15mm}\\ 
\noindent
\textbf{Acknowledgements.}  The authors are grateful to P. Baseilhac for numerous discussions. The research of L.Vinet is supported in part by a Discovery Grant from the Natural Science and Engineering Research Council (NSERC) of Canada. P.-A. Bernard and D. Shaaban Kabakibo hold a NSERC graduate scholarship.

\appendix

\section{Bethe ansatz from the lowest weight \label{app:lw}}

The commutation relation of the shifted transfer matrix with $C(v,n)$ is given by 
\begin{eqnarray}
 t(u,n) C(v,n) - C(v,n) t(u,n-1) \!\! &=&\!\! 4 C(v,n) \left( \frac{v(u^2-1)}{(u-v)(uv-1)}\left( \widetilde D(u)-\widetilde A(u) + 4\frac{u^2+1}{u^2-1}   \right) -2 \right)\qquad \\
 &&\hspace{-1cm} -4 \frac{v}{u}C(u,n)     \left( \frac{u(v^2-1)}{(u-v)(uv-1)}\left( \widetilde D(v)-\widetilde A(v) + 4\frac{v^2+1}{v^2-1}   \right) +4n \right). \nonumber
\end{eqnarray}
Let $\overline{\omega}_s$ (with $s \in \mathbb{N}/2$) denote the lowest weight vector of the spin $s$ representation of $\mathfrak{su}(2)$ which satisfies 
\begin{eqnarray}
 \widetilde J_3\overline{\omega}_s =-s \overline{\omega}_s,\qquad \widetilde J_-\overline{\omega}_s=0.
\end{eqnarray}
From the explicit form of the operators, one deduces that 
\begin{eqnarray}
 &&\hspace{-1.5cm}\widetilde A(u)\overline{\omega}_s =\overline{\alpha}(u)\overline{\omega}_s=\Big( \rho_3  - \frac{2as(u^2-1)}{(au-1)(a-u)}    -\frac{i(a^2-1)(u^2+1)\rho_1}{2a(u^2-1)} -\frac{(a-u)^2+(au-1)^2}{2a(u^2-1)} \rho_2  \Big)\overline{\omega}_s ,\\
 &&\hspace{-1.5cm}\widetilde D(u)\overline{\omega}_s =\overline{\delta}(u)\overline{\omega}_s= \Big(  \rho_3 + \frac{2as(u^2-1)}{(au-1)(a-u)}  +\frac{i(a^2-1)(u^2+1)\rho_1}{2a(u^2-1)} +\frac{(a-u)^2+(au-1)^2}{2a(u^2-1)} \rho_2  \Big)\overline{\omega}_s ,\\
 &&\hspace{-1.5cm} B(u,n) \overline{\omega}_s =\frac{1}{u} \beta_n \overline{\omega}_s=  \frac{4}{u}\left( - s -n+\frac{1}{2}+\frac{i(a^2-1)\rho_1}{4a}+\frac{(a^2+1)\rho_2}{4a}+\frac{i\rho_3}{2}   \right)\overline{\omega}_s. \label{eq:beta}
\end{eqnarray}
We construct as follows the Bethe vectors, that depend on the parameters $\overline{\boldsymbol{z}}=\{\overline{z}_1, \overline{z}_2,\dots, \overline{z}_M\}$,
\begin{equation}
 \overline{\mathbb{V}}(\overline{\boldsymbol{z}})=C(\overline{z}_1,0)C(\overline{z}_2,-1)\dots C(\overline{z}_M,-M+1) \overline{\omega}_s .
\end{equation}
Due to relation \eqref{eq:com2}, the entries of the vector $\overline{\mathbb{V}}(\overline{\boldsymbol{z}})$ do not depend on the order of the parameters $\overline{z}_i$.
The action of the transfer matrix on the Bethe vector $\overline{\mathbb{V}}(\overline{\boldsymbol{z}})$ is given by 
\begin{eqnarray}
 t(u) \overline{\mathbb{V}}(\overline{\boldsymbol{z}})&=& \frac{2}{u} \beta_{-M}  C(\overline z_1,0)C(\overline z_2,-1)\dots C(\overline z_{M},-M+1)C(u,-M) \overline{\omega}_s \nonumber\\
 &&+ \overline{\mathcal{W}}(u,\overline{\boldsymbol{z}}) \overline{\mathbb{V}}(\overline{\boldsymbol{z}}) \nonumber\\
 &&-\sum_{k=1}^M \frac{16\overline{z}_k(\overline{z}_k^2-1)}{(u-\overline{z}_k)(u\overline{z}_k-1)}   \overline{\mathcal{U}}_k(\mathcal{\boldsymbol{z}}) 
 \overline{\mathbb{V}}(\overline{\boldsymbol{z}}_k,u) \label{eq:tob}
\end{eqnarray}
where $\overline{\mathbb{V}}(\mathcal{\boldsymbol{z}}_k,u)=C(\overline{z}_1,0) \dots C(\overline{z}_{k-1},-k+2) C(u,-k+1) C(\overline{z}_{k+1},-k)\dots C(\overline{z}_M,-M+1) \overline{\omega}_s$ and 
\begin{eqnarray}
&&\hspace{-1.5cm}\overline{\mathcal{W}}(u,\overline{\boldsymbol{z}}) = \overline{\lambda}(u) +
\sum_{k=1}^M \frac{16\overline{z}_k(u^2-1)}{(u-\overline{z}_k)(u\overline{z}_k-1)}\left( \frac{\overline{\delta}(u)-\overline{\alpha}(u)}{4} +\frac{u^2+1}{u^2-1}
+\sum_{\genfrac{}{}{0pt}{2}{p=1}{p\neq k}}^M\frac{(u^2-1)\overline{z}_k\overline{z}_p}{u(\overline{z}_k-\overline{z}_p)(\overline{z}_k\overline{z}_p-1)}  \right)\!,\label{eq:Wabb}\\
&&\hspace{-1.5cm}  \overline{\lambda}(u)= \overline{\alpha}^2(u)+\overline{\delta}^2(u)+2-2u(\overline{\alpha}'(u)-\overline{\delta}'(u))
  -2\frac{u^2+1}{u^2-1}(\overline{\alpha}(u)-\overline{\delta}(u)),\\
&& \hspace{-1.5cm} \overline{\mathcal{U}}_k(\overline{\boldsymbol{z}})=\frac{\overline{\delta}(z_k)-\overline{\alpha}(z_k)}{4}
  +\frac{\overline{z}_k^2+1}{\overline{z}^2_k-1}
  +\sum_{\genfrac{}{}{0pt}{2}{p=1}{p\neq k}}^M \frac{\overline{z}_p(\overline{z}_k^2-1)}{(\overline{z}_k-\overline{z}_p)(\overline{z}_k\overline{z}_p-1)}.
\end{eqnarray}


\begin{thebibliography}{99}

\bibitem{ACDE} M.C. Albanese, M. Christandl, N. Datta, and A. Ekert,
\textsl{Mirror inversion of quantum states in linear registers,}
Phys. Rev. Lett. 93 (2004) 230502 and \texttt{arXiv:quant-ph/0405029}.

\bibitem{BellIII} J. Avan, S. Belliard, N. Grosjean, and R.A. Pimenta,
\textsl{Modified algebraic Bethe ansatz for XXZ chain on the segment - III - Proof,}
Nucl. Phys. B899 (2015) 229-246 and \texttt{arXiv:1506.02147}.

\bibitem{BBC}P. Baseilhac, S. Belliard, and N. Crampe,
\textsl{FRT presentation of the Onsager algebras,}
Lett. Math. Phys. 108 (2018) 2189 and \texttt{arXiv:1709.08555}.

\bibitem{BC1}P. Baseilhac, and N. Crampe,
\textsl{FRT presentation of classical Askey-Wilson algebras,}
Lett. Math. Phys. 109 (2019) 2187 and \texttt{arXiv:1806.07232}. 

\bibitem{Bas} P. Baseilhac, and R.A. Pimenta, 
\textsl{Diagonalization of the Heun--Askey--Wilson operator, Leonard pairs and the algebraic Bethe ansatz,}
Nucl. Phys. B949 (2019) 114824 and \texttt{arXiv:1909.02464}.

\bibitem{BTVZ} P. Baseilhac, S. Tsujimoto, L. Vinet, and A. Zhedanov, 
\textsl{The Heun--Askey--Wilson algebra and the Heun operator of Askey--Wilson type,}
Annales Henri Poincar\'e 20 (2019) 3091-3112 and \texttt{arXiv:1811.11407}.

\bibitem{BC}S. Belliard, and N. Crampe,
\textsl{Heisenberg XXX Model with General Boundaries: Eigenvectors from Algebraic Bethe Ansatz,}
SIGMA 9 (2013) 072 and \texttt{arXiv:1309.6165}.

\bibitem{BellII}S. Belliard, and R.A. Pimenta,
\textsl{Modified algebraic Bethe ansatz for XXZ chain on the segment - II - general cases,}
Nucl. Phys. B894 (2015) 527-552 and \texttt{arXiv:1412.7511}.

\bibitem{BCTVZ} G. Bergeron, N. Crampe, S. Tsujimoto, L. Vinet, and A. Zhedanov,
\textsl{The Heun--Racah and Heun--Bannai--Ito algebras,}
J. Math. Phys. 61 (2020) 081701 and \texttt{arXiv:2003.09558}.

\bibitem{CYSW}
 J. Cao,  W.L. Yang, K. Shi, and Y.  Wang, \textsl{Off-diagonal  Bethe  ansatz  solution  of  the  XXX  spin-chain  with arbitrary boundary conditions},
 Nucl. Phys. B875 (2013) 152-165 and \texttt{arXiv:1306.1742}. 


\bibitem{CDD}M. Christandl, N. Datta, T.C. Dorlas, A. Ekert, A. Kay, and A.J. Landahl,
\textsl{Perfect transfer of arbitrary states in quantum spin networks,}
 Phys. Rev. A032312 (2005) 71 and \texttt{arXiv:quant-ph/0411020}.
 
\bibitem{CraTASEP} N. Crampe,
\textsl{Algebraic Bethe ansatz for the totally asymmetric simple exclusion process with boundaries,}
J. Phys. A: Math. Theor. 48 (2015) 08FT01 and \texttt{arXiv:1411.7954}. 

\bibitem{Cra} N. Crampe,
\textsl{Algebraic Bethe ansatz for the XXZ Gaudin models with generic boundary,}
SIGMA 13 (2017) 094 and \texttt{arXiv:1710.08490}.

\bibitem{CKV} N. Crampe, K. Guo, and L. Vinet, 
\textsl{Entanglement of free Fermions on Hadamard graphs,}
Nucl. Phys. B960 (2020) 115176 
and \texttt{arXiv:2008.04925}.

\bibitem{CN}N. Crampe, and R.I. Nepomechie,
\textsl{Equivalent T-Q relations and exact results for the open TASEP,}
J. Stat. Mech. (2018) 103105 and \texttt{arXiv:1806.07748}.

\bibitem{CNV}N. Crampe, R.I. Nepomechie, and L.Vinet,
\textsl{Free-Fermion entanglement and orthogonal polynomials},
J. Stat. Mech. (2019) 093101 and \texttt{arXiv:1907.00044}.

\bibitem{CNV2}N. Crampe, R.I. Nepomechie, and L.Vinet,
\textsl{Entanglement in Fermionic Chains and Bispectrality},
Roman Jackiw 80th Birthday Festschrift (World Scientific, 2020) and \texttt{arXiv:2001.10576}.

\bibitem{CPV} N. Crampe, L. Poulain d'Andecy, and L. Vinet,
\textsl{A Calabi--Yau algebra with $E_6$ symmetry and the
Clebsch--Gordan series of $sl(3)$,}
\texttt{arXiv:2005.13444}.

\bibitem{CVZ} N. Crampe, L. Vinet, and A. Zhedanov,
\textsl{Heun algebras of Lie type,}
Proc. Amer. Math. Soc. 148 (2020) 1079-1094  and \texttt{arXiv:1904.10643}.

\bibitem{EP}
V. Eisler, and I. Peschel, 
\textsl{Properties of the entanglement Hamiltonian for finite free-fermion chains},
J. Stat. Mech. 10 (2018) 104001 and \texttt{arXiv:1805.00078}.

 \bibitem{GVZ} F.A. Gr\"unbaum, L. Vinet, and A. Zhedanov,  
 \textsl{Algebraic Heun Operator and Band-Time Limiting,}
 Comm.  Math. Phys. 364 (2018) 1041-1068 and \texttt{arXiv:1711.07862}.

 \bibitem{GVZ2}  F.A. Gr\"unbaum, L. Vinet, and A. Zhedanov,  
 \textsl{Tridiagonalization and the Heun equation,}
 J. Math. Phys. 58 (2017) 1-17 and \texttt{arXiv:1602.04840}.
  
  \bibitem{ISL}
  M.E.H. Ismail, S.S. Lin, and S.S. Roan, 
  \textsl{Bethe Ansatz Equations of XXZ Model and q-Sturm-Liouville Problems,}
  \texttt{arXiv:math-ph/0407033}.
  
  
  \bibitem{KS} R. Koekoek, and R.F. Swarttouw,
\textsl{The Askey-scheme of hypergeometric orthogonal polynomials and its $q$-analogue,}
\texttt{arXiv:math/9602214}.

  \bibitem{PW_Heun} J. Patera, and P.  Winternitz, 
{\sl A new basis for the representations of the rotation group. Lam\'e and Heun polynomials}, 
J. Math. Phys. {\bf 14} (1973) 1130-1139.

   \bibitem{Skl} E.K. Sklyanin, 
 \textsl{Boundary conditions for integrable quantum systems}, 
 J. Phys. A: Math. Gen.21 (1988) 2375-2389.
  
 \bibitem{STF} E.K. Sklyanin, L.A. Takhtadzhyan, and L.D. Faddeev, 
 \textsl{Quantum inverse problem method. I}, 
 Theor. Math. Phys. 40 (1979) 688--706.
 
 \bibitem{Skr} T. Skrypnyk,
\textsl{Generalized Gaudin systems in a magnetic field and non-skew-symmetric $r$-matrices,}
 J. Phys. A: Math. Theor. 40 (2007) 13337.
 
 \bibitem{T} A.V. Turbiner, 
  {\sl The Heun operator as a Hamiltonian,}
J. Phys. A: Math. Theor. 49 (2016) 26LT01 and \texttt{arXiv:1603.02053}. 

\bibitem{VZ}L. Vinet, and A. Zhedanov,
\textsl{The Heun operator of Hahn type,}
Proc. Amer. Math. Soc. 147 (2019) 2987-2998 and \texttt{arXiv:1808.00153}. 

\bibitem{WZ}P.B. Wiegmann, and A.V. Zabrodin,
\textsl{Algebraization of difference eigenvalue equations related to $Uq(sl_2)$},
Nucl. Phys. B 451 (1995) 699 and \texttt{cond-mat/9501129}.

\end{thebibliography}
\end{document}